\documentclass[12pt]{article}
\usepackage{amssymb,tabularx,multirow,amsmath,natbib, epsfig, threeparttable, amstext}
\usepackage{setspace}
\usepackage{verbatim}
\usepackage{times}
\usepackage{fancyhdr}
\usepackage{relsize}
\usepackage{hyperref}
\usepackage{booktabs}

\newcommand{\mbf}[1]{\mathbf{#1}}
\newcommand{\mbv}[1]{\mbox{\boldmath$#1$\unboldmath}}

\usepackage{setspace}
\RequirePackage[vmarginratio=1:2, outer=30mm, inner=26mm, lines=26]{geometry}
\usepackage{graphicx}
\usepackage{indentfirst}
\usepackage{caption}
\usepackage{subcaption}
\usepackage{natbib}
\usepackage{mathrsfs} 
\usepackage{floatrow}
\usepackage{framed}

\newcommand{\bes}{\begin{slide*}{}}
\newcommand{\es}{\end{slide*}}

\newcommand{\bitem}{\begin{itemize}}
\newcommand{\eitem}{\end{itemize}}
\newcommand{\blist}{\begin{list}{$\bullet$}{}}
\newcommand{\elist}{\end{list}}
\newcommand{\barray}{\begin{eqnarray*}}
\newcommand{\earray}{\end{eqnarray*}}
\newcommand{\bright}{\begin{flushright}}
\newcommand{\eright}{\end{flushright}}

\usepackage{booktabs}

\newcommand{\be}{\begin{eqnarray}}
\newcommand{\ee}{\end{eqnarray}}
\newcommand{\ba}{\begin{eqnarray*}}
\newcommand{\ea}{\end{eqnarray*}}

\def\boldfacefake #1{%
	\hbox{%
		\mathsurround=0pt
		\hbox to 0.25pt{$#1$\hss}%
		\hbox to 0.25pt{$#1$\hss}%
		\hbox {$#1$}%
	}%
}
\makeatletter
\def\@seccntformat#1{%
  \expandafter\ifx\csname c@#1\endcsname\c@subsection\else
  \csname the#1\endcsname\quad
  \fi}
  
\makeatother

\newcommand{\expect}{\mbox{\rm I\kern-.20em E}}
\newcommand{\reals}{\mbox{\rm I\kern-.20em R}}
\newcommand{\sreals}{\mbox{\small \rm I\kern-.20em R}}

\def\ba{\mathbf{a}}

\def\br{\mathbf{r}}
\def\bs{\mathbf{s}}

\def\bA{\mathbf{A}}
\def\bB{\mathbf{B}}

\def\bD{\mathbf{D}}
\def\bE{\mathbf{E}}
\def\bF{\mathbf{F}}

\def\bH{\mathbf{H}}
\def\bI{\mathbf{I}}

\def\bU{\mathbf{U}}
\def\bR{\mathbf{R}}

\def\bV{\mathbf{V}}

\def\bW{\mathbf{W}}

\def\bY{\mathbf{Y}}

\usepackage{color}
\usepackage{bm}
\usepackage{multirow}
\usepackage{amsmath}
\usepackage{amsfonts}
\usepackage{setspace}
\PassOptionsToPackage{round}{natbib}
\usepackage{natbib}
\usepackage{lineno}

\usepackage{setspace}
\begin{document}
\bibliographystyle{asa}
\doublespacing
%\begin{titlepage}

\title{A Hierarchical Spatio-Temporal Analog Forecasting Model for Count Data}
\date{\vspace{-5ex}}
\author{Patrick L. McDermott\thanks{Correspondence to: P.L. McDermott, Department of Statistics, University of Missouri,  146 Middlebush Hall, Columbia, MO 65211 U.S.A.. E-mail: plmyt7@mail.missouri.edu} \footnotemark[2] ,\,\, Christopher K. Wikle\thanks{Department of Statistics, University of Missouri, Columbia, MO, 65211, U.S.A.},\,\, Joshua Millspaugh\thanks{Wildlife Biology Program, University of Montana, Missoula, MT, 59812, U.S.A.}}

\maketitle
\section*{Summary}
\begin{enumerate}
\item Analog forecasting has been successful at producing robust forecasts for a variety of ecological and physical processes. Analog forecasting is a mechanism-free nonlinear method that forecasts a system forward in time by examining how past states deemed similar to the current state moved forward. Previous work on analog forecasting has typically been presented in an empirical or heuristic context, as opposed to a formal statistical context.
\item The model presented here extends the model-based analog method of \cite{analog2016} by placing analog forecasting within a fully hierarchical statistical framework. In particular, a Bayesian hierarchical spatial-temporal Poisson analog forecasting model is formulated.  
% \item Various nonlinear dimension reduction techniques are explored to identify finding robust analogs. Nonnegative matrix factorization (NMF) is used to efficiently reduce the dimension of the count-valued response.
\item In comparison to a Poisson Bayesian hierarchical model with a latent dynamical spatio-temporal process, the hierarchical analog model consistently produced more accurate forecasts. By using a Bayesian approach, the hierarchical analog model is able to quantify rigorously the uncertainty associated with forecasts.
 \item Forecasting waterfowl settling patterns in the northwestern United States and Canada is conducted by applying the hierarchical analog model to a breeding population survey dataset. Sea Surface Temperature (SST) in the Pacific ocean is used to help identify potential analogs for the waterfowl settling patterns. 
\end{enumerate}
\textbf{Keywords:} Nonlinear forecasting; hierarchical Bayesian models; ecological forecasting; waterfowl settling patterns

%\linenumbers
%\end{titlepage}
\setcounter{page}{2}

\section*{Introduction}

Contemporary issues in natural resource management such as climate change rely increasingly on quantitative forecasts at time scales ranging from seasonal to decadal \citep[e.g.,][]{lebrun2016assessing}.  There are great challenges when making such forecasts in a rapidly changing environment.  One of the most important challenges to policy and management is to quantify the uncertainty of the forecasts \citep[e.g.,][and references therein]{clark2001ecological,conroy2011conservation}.  There are many potential issues with quantifying uncertainty, related to the characterization of uncertainties in data, mechanistic processes, and interactions across biological and physical systems \citep[e.g.,][]{oliver2015pitfalls}.  Perhaps surprisingly, in many cases, the best forecast models rely on non-parametric and ``mechanism-free'' specifications \citep[e.g.,][]{perretti2013model, ward2014complexity}.  Bayesian models in general, and Bayeisan hierarchical models in particular, provide a comprehensive modeling framework which account for multiple sources of uncertainty in ecological models \citep[e.g.,][to name a few]{wikle2003,royle2008hierarchical,cressie2009}; for a historical overview see \cite{ellison2004bayesian}.  To date, there have been few attempts to cast ``mechanism-free'' models within the Bayesian framework \citep{analog2016}.

Quantifying uncertainty for spatial-temporal ecological processes is complicated because the evolution of these processes over time is often nonlinear. One mechanism-free solution to the spatio-temporal forecasting problem is known as ``analog forecasting'' \citep[e.g.,][]{lorenz1969atmospheric}. Analog forecasting uses past states of a system that are similar to the current state and then assumes that the current state of the system will evolve in a manner similar to how the identified past states evolved. Analog forecasting is appealing for dynamical processes governed by some underlying, but unspecified, deterministic law.  Specifically, analog forecasting leverages the predictability in these types of systems by finding past trajectories similar to the current trajectory of the system. 

Much of the current development of spatio-temporal analog methods utilizes the idea of embedding a dynamical system in time, similar to the simplex prediction method outlined in \cite{sug} for univariate time series.  Indeed, the \cite{sug} approach was one of the first practical methods to introduce the idea of embedding a dynamical system in the context of nonlinear forecasting. Their methods utilized the state-space dynamical system reconstruction theory of \cite{Takens}. In complicated dynamical systems, one rarely observes all of the state variables. State-space reconstruction allows one to reconstruct a dynamical system with only a subset of the state variables, by considering those state variables at multiple lags in the past. As dynamical systems evolve in time they tend to revisit previous paths in the phase space, where these paths live on some low-dimensional manifold of the entire space (i.e., the attractor). Thus, through the use of state-space reconstruction one can recover features of past dynamical paths along the attractor.  \cite{sug} recognized the utility of state-space reconstruction within the context of nonlinear forecasting. In particular, they showed how embedding vectors, created by lagging past states of a system (historical data) in time \citep[e.g., see Chapter 3 of][]{CandW2011}, could be utilized to find robust analogs for the current state of the system. This remarkably simple forecasting method has proved successful in a multitude of time series applications \citep[e.g.,][]{sug2012,perretti2013model,Zhao} .

Mechanism-free and analog methods traditionally have relied on non-parametric and/or heuristic approaches that did not include a formal probabilistic error structure \citep[although, see][]{Tippett}. Modern non-parametric analog methods require choice of the embedding dimension of the analogs, the number of past analogs to consider, and weights for those analogs.  All of these choices can significantly impact the analog forecast. For example, the question of how many past analogs to use can be thought of as a k-nearest neighbor problem, where the neighborhood consists of the analogs most similar to the current state of the system. Given the number of ``neighboring" analogs, a kernel defined by a smoothing parameter is typically used to determine the weights \citetext{e.g., \citealp{Zhao}; \citealp{analog2016}}. However, previous analog forecasting implementations have employed either some heuristic method that does not explicitly account for uncertainty associated with the choice, or a multidimensional cross-validation search \citep[e.g.,][]{econGNP}, to choose these values. The Bayesian framework described in \cite{analog2016} allows for both the estimation and incorporation of model averaging over the various parameters in the analog model, thereby accounting for the uncertainty induced by their selection.

Once framed within the context of Bayesian modeling, analog forecasting can be placed within the rich class of models available in the space-time hierarchical Bayesian framework \citep[e.g.,][]{CandW2011,wikle2015modern}, which allows for robust quantification of uncertainity. We present here a hierarchical analog forecasting model that extends the model developed in \cite{analog2016} to include a formal non-Gaussian data model -- specifically, a Poisson model to accommodate count data.  This is the first analog method that  accounts explicitly for non-Gaussian data within a statistical framework. The model is applied to the problem of producing one year-ahead forecasts of waterfowl settling patterns given the state of the Pacific ocean sea surface temperature (SST).  Because spatio-temporal analog forecasting can quickly become prohibitive for high-dimensional processes, we introduce an approach for spatio-temporal dimension reduction of count data known as nonnegative matrix factorization \citep{nmf2001}.  

%The paper proceeds with a description of the waterfowl settling pattern forecasting problem, followed by a description of the Breeding Population Survey and SST data sets.  We then describe the importance of dimension reduction for high-dimensional spatio-temporal data with emphasis on nonnegative matrix factorization for count data, and various traditional decompositions for continuous valued data.  The Bayesian hierarchical analog model is then described in detail, followed by results to the application to the waterfowl settling forecast problem, and a brief discussion.

\section*{Materials and Methods}
\subsection{Waterfowl and Sea Surface Temperature Data}

Migratory waterfowl settling patterns, productivity, and survival have been shown to depend strongly on climate-related habitat conditions \citep[e.g.,][]{hansen1964emigration,herter2012bird,feldman2015does}.  It is known that changes to habitat conditions can lead to more flexible settling patterns along a lattitudinal gradient that can mitigate site philopatry, and possibly decrease productivity or recruitment \citep[e.g.,][]{johnson1988determinants,karanth2014latitudinal,becker2015search}.  Given the well-known relationships between Pacific ocean (particularly the tropical ocean) SSTs and North American climate conditions \citep[e.g.,][]{philander1990} and the potential for these conditions to affect waterfowl settling patterns \citep{sorenson1998potential}, it is reasonable to use Pacific ocean SST as a proxy for future habitat conditions. In addition, the impact of Pacific SSTs is typically nonlinear \citep{hoerling1997nino}, suggesting nonlinear evolution models are appropriate. Although others \citep[e.g.,][]{wu2013hierarchical} have successfully forecast Mallard duck ({\it Anas platyrhyncho}) settling patterns using a drought severity index, we provide a one-year forecast given the Pacific SSTs through the previous May.

Since 1955 the U.S. Fisheries and Wildlife Service (USFWS) and Canadian Wildlife Service (CWS) have jointly conducted a Breeding Population Survey (BPS) in the northern United States and Canada. Each spring (mid to late May) crews consisting of one pilot and one observer fly transect lines and record counts of various waterfowl species. For selected areas, ground crews also record counts. Each 400 m wide transect is divided into a series of segments measuring 29 km in length and an entire survey covers approximately 2.1 square kilometers. The analysis conducted here consists of the 1,067 locations between $96^{\circ}-115^{\circ}$W longitude and $43^{\circ}-54^{\circ}$N latitude from 1970 through 2014. The majority of survey locations north of $54^{\circ}$ latitude have little temporal variability, with zero counts in most years and are not considered. Although the BPS survey records counts for several species, we focus on raw indicator pair counts (i.e., counts of paired ducks and lone drakes) for mallards. The raw indicated pair counts are publicly available through the FWS Division of Migratory Management (\url{https://migbirdapps.fws.gov/}). 

Monthly SST from 1970-2014 were obtained from the publicly available National Oceanic and Atmospheric Administration (NOAA) Extended Reconstruction Sea Surface Temperature (ERSST) data (\url{http://www.esrl.noaa.gov/psd/}). A subset of 3,132 locations from the ERSST data, between $30.5^{\circ}$S-$60.5^{\circ}$N latitude and $123.5^{\circ}$E-$290.5^{\circ}$E longitude with a spatial resolution of $2^{\circ} \times 2^{\circ} $, form the SST data. We follow the common procedure from the climate science literature by creating anomalies through the subtraction of location specific monthly means calculated from a climatological average spanning the period 1970-1999 \citep[e.g.,][]{wilks2011statistical}.

\subsection{Spatio-Temporal Variables}
Let $Y_{t}(\bs_i)$ be a component of a dynamical system at time $t$ with spatial locations $\{\bs_i$, $i=1,\dots,n_y\}$. Suppose we have access to data from the system for time periods $\{t=1,\dots,T\}$. The set of data at all $n_y$ locations for time period $t$ is defined as, $\bY_t \equiv (Y_t(\bs_1),\ldots,Y_t(\bs_{n_y}))'$. Here, we consider count valued data for $\bY_t$. Further, we consider the use of some spatio-temporal forcing (predictor) variable, defined as, $\mbv{x}_{t'}=(x_{t'}(\mbv{r}_1),\dots,x_{t'}(\mbv{r}_{n_x}))'$, for spatial locations, $\{\br_1,\dots,\br_{n_x}\}$ and time $t'$, to help forecast the process of interest (i.e.,  $\bY_t $). Note that the time indices $t$ and $t'$ are separated by $\tau$ period(s) (i.e.,  $\tau=0,1,2,\dots$), with potentially different time scales. As discussed in more detail below, in our application, $\tau$ represents the number of periods the response variable is forecasted into the future.  Thus, the goal here is to forecast the value of ${\mbf Y}_{T+\tau}$ given values of ${\mbf Y}_t$ for $t \leq T$ and for ${\mbf x}_{t'}$ for $t' \leq T$.  This is done by weighting the past values of ${\mbf Y}_t$ based on how well corresponding past sequences of ${\mbf x}_{t'}$ match the most recent sequence of ${\mbf x}_{t'}$ (i.e., the most recent sequence up to time $T$), as described below.

Many spatio-temporal dynamical processes can be challenging to model due to the high-dimensional nature of the spatial component. Both the BPS waterfowl settling pattern data and SST data described above can be considered high-dimensional. To efficiently model such spatio-temporal processes, some form of dimension reduction is usually performed \citep[e.g., see Chapter 7 of][]{CandW2011}. Common methods such as empirical orthogonal functions (EOFs) are not ideal for non-continuous responses such as count data because it is difficult to impose constraints (e.g., such as non-negativity). Although more general ordination methods such as principal coordinate analysis and multidimensional scaling can be useful for non-continous data \citep[e.g.,][]{ellison2004primer}, these methods also do not guarantee, in general, that after dimension reduction and projection back into physical space, that the resulting process has the same support as the original data.

%Suppose $n_y$ (the number of spatial locations for the response variable) is very large, in this case one may project $\bY_t$ onto a lower dimensional set of basis functions. To efficiently model $\bY_t$, a set of $n_\beta$ spatial basis functions, $\{\mbv{\psi}_j, \ j=1,\dots,n_\beta\}$, are calculated, such that $n_\beta<<n_y$. The collection of these basis functions, ${\mbv \Psi }\equiv [\psi_1,\dots,\psi_{n_\beta}]$, form a $n_y \times n_\beta$ basis function matrix. At time period $t$, a $n_\beta$-dimensional projection vector can be calculated by $\widetilde{\mbv{\beta}}_t=(\mbv{\Psi}'\mbv{\Psi})^{-1}\mbv{\Psi}'\bY_t$. Assume we choose EOFs as our basis functions to transform back to physical space, one simply looks at $\mbv{\Psi}\widetilde{\mbv{\beta}}_t$. Recall, we assume $\bY_t$ is count data, in this situation there is no guarantee this transformation will result in nonnegative values. It may be the case, especially for non-continuous data, that this transformation back to physical space is poor for even moderately large values of $n_\beta$. There are a number of situations in which the quality of this transformation has little effect on the overall model. In the hierarchical model presented below, we estimate these projection coefficients (i.e. $\widetilde{\mbv{\beta}}_t$), before transforming back to physical space to make forecasts, thus the necessity for quality transformations back to physical space.

\subsection{Response Vector Dimension Reduction}

Consider the case where we have $n_y$ spatial locations and the $n_y$-dimensional response vector at time $t$, $\bY_t$. We seek a $n_\beta$-dimensional expansion coefficient vector, $\mbv{\beta}_t$, associated with a set of $n_\beta$ basis functions $\{\mbv{\psi}_j, \ j=1,\dots,n_\beta\}$, where $\mbv{\psi}_j\equiv (\psi_j(\bs_1),\dots,\psi_j (\bs_{n_y}))'$. In particular, we seek a reduced dimension representation such that $n_\beta<<n_y$. When considering a linear basis expansion, then, we seek $\bY_t \approx  \mbv{\Psi}\mbv{\beta}_t$, where ${\mbv \Psi }\equiv [\psi_1,\dots,\psi_{n_\beta}]$ is a $n_y \times n_\beta$ matrix. Then, the ordinary least squares estimate of the expansion coefficients is $\widetilde{\mbv{\beta}}_t=(\mbv{\Psi}'\mbv{\Psi})^{-1}\mbv{\Psi}'\bY_t$, assuming $(\mbv{\Psi}'\mbv{\Psi})$ is invertible. In situations where $\mbv{\Psi}$ is orthogonal, this simplifies to $\widetilde{\mbv{\beta}}_t=\mbv{\Psi}'\bY_t$. As an example, $\mbv{\Psi}$ derived from the scaled left singular vectors of a full data matrix, $\bY\equiv [\bY_1,\dots,\bY_T]$, are the EOF basis functions, and are orthogonal. A reduced rank representation of the response vectors in phase space is given by $\widetilde{\bY}_t=\mbv{\Psi}\widetilde{\mbv{\beta}}_t$. Typically, one then considers the expansion coefficients, $\widetilde{\mbv{\beta}}_t$, as the time-varying variable of interest.

When $\bY_t$ has a constrained support, as with the count data of interest here, there is no guarantee that this back transformation ($\widetilde{\bY}_t=\mbv{\Psi}\widetilde{\mbv{\beta}}_t$) will result in appropriate support for the elements of $\widetilde{\bY}_t$ (e.g., non-negative values). This issue can be important in some applications, such as the analog forecasting problem of interest here, as we specify the  $\mbv{\beta}_t$'s in a hierarchical model and require non-negative values upon transformation back to physical space.

We employ nonnegative matrix factorization (NMF) \citep[e.g.,][]{nmf2001} to enforce non-negativity in the dimension reduction of the count data matrix. Given the $n_y \times T$ data matrix $\bY$, NMF gives:
\begin{equation}
\label{NMF}
\bY\approx \mbv{\Psi}\bB  \hspace{1cm} \mbv{\Psi}\geq0, \ \bB\geq0 ,
\end{equation}
where ${\mbv \Psi }$ is a $n_y \times n_\beta$ basis function matrix and the $n_\beta \times T$ matrix $\bB\equiv [{\mbv \beta}_1,\dots,{\mbv \beta}_T]$ contains (random) projection coefficients. In reference to (\ref{NMF}), the notation $\bW\geq0 $ for some matrix $\bW$, implies that each element of $\bW$ is nonnegative. NMF has been applied in a variety of disciplines because of its ability to provide efficient dimension reduction while creating nonnegative basis functions. A number of different algorithms to conduct NMF have been proposed in the literature \citep[e.g.,][]{Berry2007}, all with the goal of solving the following minimization problem:
\begin{equation}
\displaystyle\min_{\mbv{\Psi},\bB\geq 0 } \ D(\bY,\mbv{\Psi},\bB)+R(\mbv{\Psi},\bB) \ ,
\end{equation}
where $D(\bY,\mbv{\Psi},\bB)$ is a loss function and $R(\mbv{\Psi},\bB)$ is some regularization function. Unfortunately, these NMF algorithms do not produce a unique factorization. Instead, they converge to a local minimum, thus producing different factorizations for different starting values \citep[e.g.,][]{Boutsidis}. To alleviate this non-uniqueness problem in our methodology, we use the Nonnegative Double Singular Value Decomposition (NNSVD) approach of \cite{{Boutsidis}} to obtain starting values. Note that NNSVD was designed to produce fast convergence for sparse data structures (i.e., when $\bY$ contains a large number of zeros, as is the case with our BPS settling pattern data). The application to follow uses the so-called {\it off-set} NMF algorithm of \cite{Badea}.

\subsection{Forcing Vector Dimension Reduction}
The purpose of the forcing variables, $\{\mbv{x}_{t'}\}$, is to identify robust analogs to help predict the response variable. Further, the success of any analog forecasting model is largely determined by its ability to find robust analogs. If $n_x$ is large, we typically must reduce the dimension of the process by using spatial basis functions, $\mbv{\Phi}\equiv[\mbv{\phi}_1,\dots,\mbv{\phi}_{n_\alpha}]$, where  $\mbv{\phi}_k=(\phi_k(\br_1),\dots,\phi_k(\br_{n_x}))'$. As with the response vector, if we assume linear projections, we can get projection coefficients by $\mbv{\alpha}_{t'}=(\mbv{\Phi}'\mbv{\Phi})^{-1}\mbv{\Phi}'\mbv{x}_{t'}$. \cite{analog2016} show that these projection coefficients can be combined to form time lagged {\it embedding matrices}. That is, let $q$ represent the number of periods lagged back in time, then for period $t$ we can define the following $n_\alpha \times q$ embedding matrix:
\begin{equation}
\label{embedMatrix}
\bA_t=[{\mbv \alpha}_{t'},{\mbv \alpha}_{t'-1},\dots, {\mbv \alpha}_{t'-(q-1)}].
\end{equation}
These embedding matrices are critical to the success of the analog forecasting model outlined below. For example, suppose we wanted to investigate if the response variable at period $t-1$ was a robust analog for the response at period $t$. One could construct an embedding matrix $\bA_t$ corresponding to period $t$ and another matrix $\bA_{t-1}$ for period $t-1$. We could assess the quality of $\bY_{t-1}$ as an analog for the response at period $t$, by examining the ``distance" between $\bA_t$ and $\bA_{t-1}$. 

The selection of basis functions to obtain $\mbv{\alpha}_{t'}$ can be flexible here and different choices of $\mbv{\Phi}$ could potentially produce different sets of analogs. For example, EOFs would be an obvious choice if linearity was assumed. However, there is scientific evidence of a nonlinear relationship between precipitation (which could potentially affect habitat conditions) and SST anomalies \citep[e.g.,][]{hoerling1997nino}, so we investigated several nonlinear dimension reduction techniques for the waterfowl settling pattern application.

\subsection{Hierarchical Analog Forecasting Model} 
%Prior to introducing the hierarchical analog forecasting model, we review briefly the mechanics of hierarchical Bayesian modeling  \citetext{for an overview in ecology see, e.g.:  \citealp{wikle2003}, \citealp{clark2007models}; \citealp{royle2008hierarchical}; \citealp{cressie2009}; \citealp{hobbs2015bayesian}}. If we let $X$ represent some latent dynamical process with observed data $Y$ and parameters $\Theta$ then we can model the process hierarchically as follows:
% \begin{equation}
% [  X, \Theta \mid Y ]\propto [Y \mid X] \ [X \mid \Theta] \ [\Theta] ,
% \end{equation}
% where the notation $[ \ \cdot \ ]$ denotes a probability distribution. In this modeling framework, outlined in \cite{Berliner} and \cite{wikle1998hierarchical}, $[Y \mid X] $ is generally referred to as the {\it data model}, and $[X \mid \Theta]$ and $ [\Theta] $ represent the {\it process model} and {\it parameter model}, respectively. Each stage can contain multiple substages \citep[e.g.,][]{cressie2009} and scientific knowledge can be incorporated into this framework along with models for both spatial and temporal dependences. The primary benefit of the approach is that it accommodates uncertainty at each modeling level, and the conditional structure simplifies model construction and eases computation.
 
We now discuss the specifics of the spatio-temporal hierarchical Bayesian analog (HBA) forecasting model for count data. All of the stages of the presented HBA model are summarized in Table \ref{table_1} below.  Since our responses $\{\bY_t: t=1,\dots,T \}$ are count valued, we model the data with a Poisson distribution conditional on a spatio-temporal intensity process as:
 \begin{equation}
 \label{data_model}
 \bY_t \mid \mbv{\lambda}_t \;  \sim \; \text{Poi}(\mbv{\lambda}_t),
 \end{equation}
 where $\{\mbv{\lambda}_t : t=1,\dots,T \}$ is the $n_y$-dimensional intensity process at locations $\{\bs_1,\dots,\bs_{n_y} \}$. Using the basis functions from the NMF approximation (\ref{NMF}), let $\mbv{\lambda}_t=\mbv{\Psi}\mbv{\beta}_t$. Recall, the NMF guarantees $\mbv{\Psi} \geq0$ and thus, for $\mbv{\lambda}_t$ to be nonnegative, the distribution for $\mbv{\beta}_t$ should have nonnegative support. If we denote the model parameters by $\widetilde{\mbv{\Theta}}$ (see below), then for period $t$, the process model on ${\mbv \beta}_t$ is given by the truncated normal distribution:
  \begin{equation}
  \label{processModel}
 {\mbv \beta}_t \mid \bB_{-t}, \widetilde{\Theta} \; \sim  \; \text{TN}_{[0,\infty)}( \ \text{max}\{ \ h( \ \bB_{-t} \ {\mbv \omega}_t,\sigma^2_\eta \ ), \ \epsilon\} \ ,\sigma^2_\eta  \ ) ,
  \end{equation}
where, for period $t$, we define $\bB_{-t}\equiv [{\mbv \beta}_1,\dots,{\mbv \beta}_{t-1},{\mbv \beta}_{t+1},\dots,{\mbv \beta}_{T}]$ as the matrix of possible analogs and ${\mbv \omega}_t=(\omega(\bA_t,\bA_1,{\mbf \theta}),\dots,\omega(\bA_t,\bA_{t-1},{\mbf \theta}),\omega(\bA_t,\bA_{t+1},{\mbf \theta}),\dots,\omega(\bA_t,\bA_T,{\mbf \theta}))'$, as the weight associated with each of the potential analogs.  Thus, a weighted prediction of the new ${\mbv \beta}_t$ is based on the linear combination of past ${\mbv \beta}_t$ values, ${\mbf B}_{-t} {\mbv \omega}_t$.  Further,  as described in \cite{cangelosi}, for a normal density left-truncated at zero, the mean is biased and this bias increases for values close to zero (which is the case for many elements of $\mbv{\beta}_t$) since the left tail of the distribution has been distorted from the truncation at zero. In equation (\ref{processModel}), $h(\ \cdot \ )$ is the bias correction function presented in \cite{cangelosi}. The need for the constant $\epsilon$ arises because as $\bB_{-t} \ {\mbv \omega}_t \to 0$, we have $h( \ \bB_{-t} \ {\mbv \omega}_t,\sigma^2_\eta \ ) \to -\infty$. Thus, $\epsilon$ is set to an arbitrarily small value for computational purposes.   
 The weights (${\mbv \omega}_t$) in (\ref{processModel}) are critical to the success of the analog forecasting model presented here. For example, during the training of the model, these weights determine how much each potential analog in $\bB_{-t}$ is weighted in order to predict ${\mbv \beta}_t$. We describe our choice of weights in the next section. %Indeed, choosing a method to determine these weights is a fundamental decision in any analog method. \cite{Zhao} detail  rigorously several possible methods for determining these weights. We follow the methodology of \cite{analog2016} that uses a Gaussian kernel with Procrustes distance metric. 
 It is important to note that, although the weights are applied to the potential analogs in a linear fashion (i.e., $ \bB_{-t} \ {\mbv \omega}_t$), the resulting prediction for ${\mbv \beta}_t$ can be considered nonlinear since the weights are determined by a nonlinear function (i.e., the Gaussian kernel).
 
\begin{table}[H]
  \begin{framed}
\captionsetup{font=footnotesize}
 \makebox[\linewidth]{\resizebox{1\linewidth}{!}{%
  \tabcolsep=.001pt%
\begin{tabular}{lc} 
 & Hierarchical Bayesian Analog Model   \\
\hline
{\bf Data model:} & $  \bY_t \mid \mbv{\lambda}_t \; \sim \; \text{Poi}(\mbv{\lambda}_t)$ \\
{\bf Process model:} & $ {\mbv \beta}_t \mid \bB_{-t}, \widetilde{\Theta} \; \sim \; \text{TN}_{[0,\infty)}( \ \text{max}\{ \ h( \ \bB_{-t} \ {\mbv \omega}_t,\sigma^2_\eta \ ), \ \epsilon\} \ ,\sigma^2_\eta  \ )$ \\
 \hspace{3mm} where & $\bB_{-t}\equiv [{\mbv \beta}_1,\dots,{\mbv \beta}_{t-1},{\mbv \beta}_{t+1},\dots,{\mbv \beta}_{T}]$ \\
& ${\mbv \omega}_t \equiv (\omega(\bA_t,\bA_1,{\mbf \theta}),\dots,\omega(\bA_t,\bA_{t-1},{\mbf \theta}),\omega(\bA_t,\bA_{t+1},{\mbf \theta}),\dots,\omega(\bA_t,\bA_T,{\mbf \theta}))'$  \\
{\bf Parameter model:} & $q \sim \text{DU}(q_{min},q_{max})$ \hspace{.2cm} $m \sim \text{DU}(m_{min},m_{max})$ \\
& $\theta_1 \sim \text{IG}(a_1,b_1)$ \hspace{.2cm} $\sigma^2_\eta \sim \text{IG}(a_2,b_2)$\\
{\bf Hyperparameters:} & $\epsilon$, $q_{min}$, $q_{max}$, $m_{min}$, $m_{max}$, $a_1$, $b_1$, $a_2$, $b_2$, $\theta_2$\\
\end{tabular}}}
\caption{Hierarchical model summary. }
\label{table_1}
\end{framed}
\end{table}

\begin{figure}[htb]
  \centering
\captionsetup{font=footnotesize}
\includegraphics[width=14cm,height=10cm]{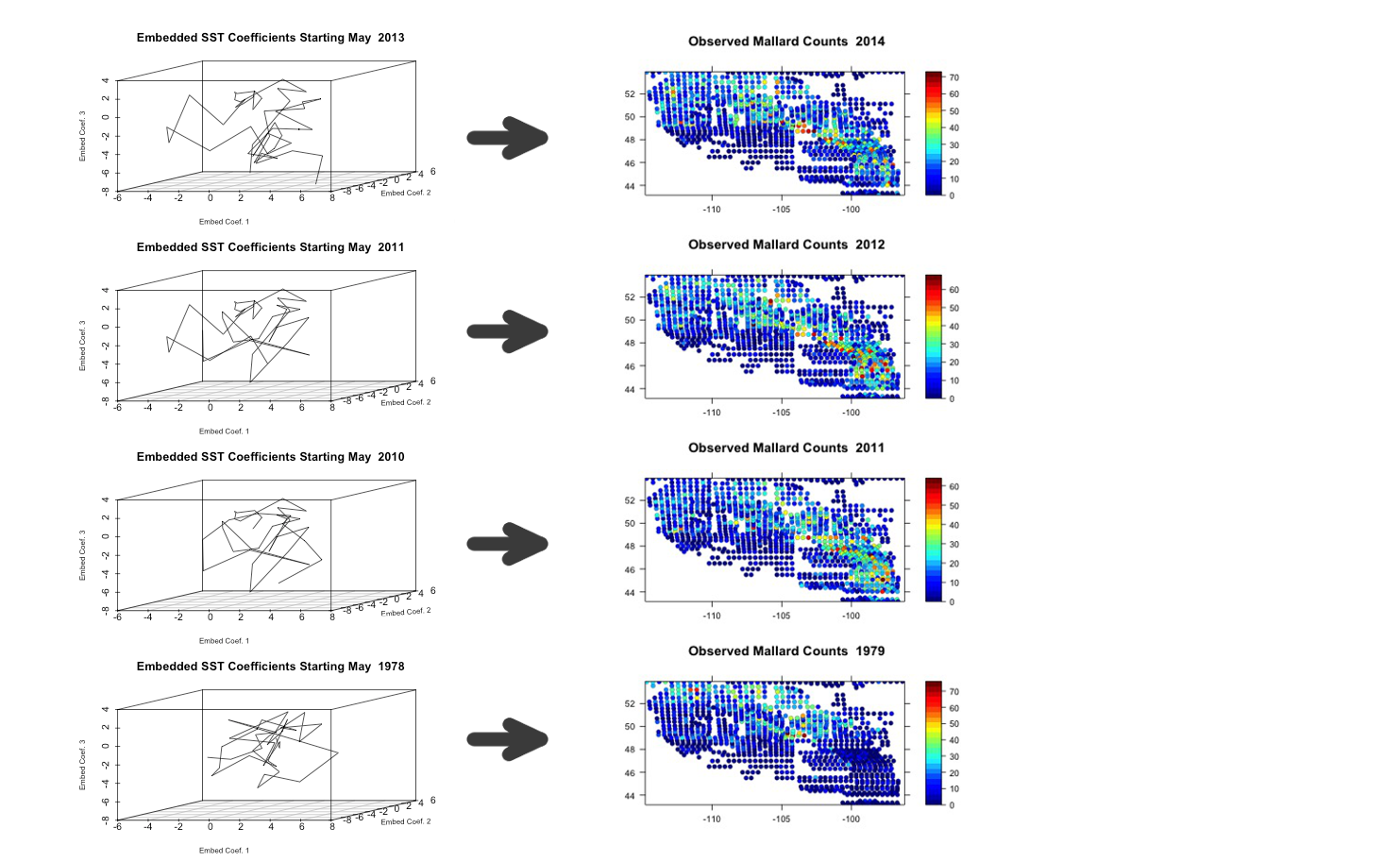}
\caption{Example illustrating analog forecasting of waterfowl counts for 2014. Attractor manifold plots on the left are examples of embedding matrices (see (\ref{embedMatrix})), where $n_\alpha=3$ and $q=50$ (months). The three  plots below the plot starting in May 2013 on the left side are examples of nearest neighbor analogs. These three neighbors are selected based on their similarity in shape (Procrustes distance) to the attractor time series for May 2013 (i.e., the initial condition for a one-year ahead forecast for May 2014). Each of the three nearest neighbors is associated with a corresponding waterfowl pattern (right column). The three waterfowl patterns for the nearest neighbors are each then appropriately weighted to form a forecast for 2014.}  
\label{Figure_1}
\end{figure}

The choice of analog weights and the analog ``neighborhood'' are closely connected and important considerations in analog forecasting.   
%Determination of how many analogs to consider is based on a nearest neighbors approach. 
Let ${\cal N}_m(\bA_t)$ denote the neighborhood of the analog $\bA_t$ for period $t$, where the number of nearest neighbors considered is represented by $m$. Defining $d( \ \cdot \ )$ as a distance metric and ${\mbf \theta}=\{\theta_1,\theta_2\}$ as a set of kernel dependent parameters, we have the following kernel weight function:
  \begin{align}
 \label{kern}
\tilde{\omega}(\bA_t,\bA_\ell,{\mbf \theta})=
    \begin{cases}
  \  \text{exp}\left(\frac{-d({\mbf A}_{t},{\mbf A}_{\ell};\theta_2)^2}{{2}\theta_1}\right)  \ , & \text{if}\ \   {\mbf A}_{\ell}\in {\cal N}_m(\bA_t)  \\
      0, & \text{if}\ \  {\mbf A}_{\ell}\notin {\cal N}_m(\bA_t)  ,  \
    \end{cases}
  \end{align}
for $\ell\neq t$, where $\theta_1$ is a kernel smoothing parameter and $\theta_2$ is a parameter associated with the distance function (see the Appendix). Let, $\omega(\bA_t,\bA_\ell,{\mbf \theta})$ be the normalized version of $\tilde{\omega}(\bA_t,\bA_\ell,{\mbf \theta})$, where the normalization is accomplished by dividing by the sum of $\tilde{\omega}(\bA_t,\bA_\ell,{\mbf \theta})$  across all $T-1$ potential analogs for period $t$. Any valid distance metric $d( \ \cdot \ )$ can be applied here; e.g., analog forecasting methods traditionally use Euclidean distance. However, analog forecasting relies on identifying analogs that not only resemble the current state of the system, but also move forward in a similar manner. For this reason, analogs that share a similar trajectory in phase space as the current trajectory of the system will produce the most successful forecasts. Procrustes distance \citep[e.g., see][]{HTF} is a multivariate distance metric that transforms a comparison object (i.e., $\bA_\ell$) to a target object (i.e., $\bA_t$), before calculating the Frobenius matrix norm between the target object and the transformed comparison object. Therefore, by defining $d({\mbf A}_{t},{\mbf A}_{\ell};\theta_2)$ as the Procrustes distance (see the Appendix for the full details, including the specification of $\theta_2$) we are able to compare the shape, and thus, the trajectory, between two embedding matrices (see Figure \ref{Figure_1} for a visual example). In the definition of $\bA_t$, we let $q$ represent the number of lagged time periods when forming $\bA_t$. Since different values of $q$ will lead to different embedding matrices, and thus potentially different analogs, we estimate $q$ and give it a discrete uniform prior such that, $q \sim \text{DU}(q_{min},q_{max})$. We also assign a discrete uniform prior to the number of neighbors parameter, $m \sim \text{DU}(m_{min},m_{max})$.  Finally, $\theta_1$ and $\sigma^2_\eta$ are both assigned inverse gamma priors, $\theta_1 \sim \text{IG}(a_1,b_1)$ and $\sigma^2_\eta \sim \text{IG}(a_2,b_2)$. 

Sampling from the posterior distribution is accomplished with Markov Chain Monte Carlo (MCMC) methods \citep[e.g.,][]{robert2004monte}. Due to the lack of conjugacy, all parameters are updated with a Metropolis-Hastings step (see the outline in the Appendix). During each iteration of the MCMC sampler, parameters are sampled using data from training periods, $t=1,\dots, T$. At this stage, all prediction is ``in-sample". For period $T+1$, out-of-sample forecasts are then drawn from the posterior prediction distribution, $\bY_{T+1}^{(\ell)} \sim \text{Poi}(\mbv{\Psi}\mbv{\beta}_{T+1}^{(\ell)})$, after each iteration, $\ell$, of the sampler. By defining,  $\bB_{T+1}^{(\ell)}=[{\mbv \beta}^{(\ell)}_1,\dots,{\mbv \beta}^{(\ell)}_{T}]$ and ${\mbv \omega}^{(\ell)}_{T+1}=(\omega(\bA_{T+1},\bA_1,{\mbf \theta}^{(\ell)}),\dots, \omega(\bA_{T+1},\bA_T,{\mbf \theta}^{(\ell)}))'$, the projection coefficients for period $T+1$ can be forecasted for the $\ell^{th}$ iteration as, $\mbv{\beta}_{T+1}^{(\ell)}= \bB_{T+1}^{(\ell)} {\mbv \omega}_{T+1}^{(\ell)}$. In this example, $\bA_{T+1}$ is the initial condition for which we seek matching past analogs. Then, from the definition of (\ref{embedMatrix}), the first element of  $\bA_{T+1}$  is $\mbv{\alpha}_{T'+1}$, which is lagged $\tau$ periods behind the forecast target time, $T+1$, thus leading to a $\tau$-period ahead forecast of $\bY_{T+1}$ (see Figure \ref{Figure_1} for an illustrative example).  

\subsection{Model Setup}

We evaluate the predictive ability of the model by considering forecasts of waterfowl counts in 2009 and 2014, while also producing hindcasts for 1999. The year 2009 was chosen due to the relative lack of correlation between the mallard counts in 2009 and the prior year. Further, we choose to consider 1999 because it was a strong La Ni\~na year, which allows us to demonstrate how the model can effectively forecast years where waterfowl patterns may change due to alternating habitat conditions. All of the data prior to the respective year is used for training 2009 and 2014, while the hindcast is implemented by training on all of the data except the counts for 1999. We make one-year ahead forecasts for all time periods by setting $\tau=12$.

We compare the forecasting skill of the HBA model with a fairly state-of-the art hierarchical Bayesian Poisson space-time model (referred to as the PST model). The PST model is comprised of a Poisson data model, $ \bY_t \mid \mbv{\lambda}_t  \sim \text{Poi}(\mbv{\lambda}_t)$, and process model defined as,  $\text{log}(\mbv{\lambda}_t)\sim  \text{Gau}( {\mbv \mu}+\mbv{\Psi}\mbv{\alpha}_t, \sigma^2_\epsilon \bI) $. Here, ${\mbv \mu}$  is a spatially indexed mean (modeled with spatial covariates), and $\mbv{\alpha}_t$ are projection coefficients formed from kernel principal component analysis (see below). The projection coefficients are modeled with a reduced rank vector-autoregressive (VAR) structure such that, $\mbv{\alpha}_t\sim \text{Gau}(\bH \mbv{\alpha}_{t-1},\mbv{\Sigma}_\gamma)$ \citep[e.g., see Chapter 7 of][]{CandW2011}. Specification of the process model for the PST model can be thought of as a linear version of the regime-dependent nonlinear model presented in \cite{wu2013hierarchical}. Comparison of the posterior predictions for the HBA and PST model is carried out by using mean squared prediction error (MSPE) and the correlation between the forecasted and observed values as in \cite{analog2016}.

The HBA model was implemented for all forecasted years with the same tuning parameters and prior distributions. Note, as $n_\beta$ increases, the NMF basis function approximation in (\ref{NMF}) generally becomes more accurate. Because there is a computational cost to using higher values of $n_\beta$, we found that $n_\beta=14$ appropriately balanced computational efficiency with the accuracy of the approximation. 

Regarding the SST basis functions, in addition to the more traditional empirical orthogonal function (EOFs; i.e., spatio-temporal principal components) linear dimension reduction, we implemented the following nonlinear dimension reduction methods: local linear embeddings \citep[e.g.,][]{lle}, diffusion maps \citep[e.g.,][]{diffusionMaps}, kernel principal component analysis (KPCA) \citep[e.g.,][]{kernelPCA}, and Laplacian eigenmaps \citep[e.g.,][]{laplacianEigen}. Our analysis found Laplacian eigenmaps to be the most helpful of these nonlinear methods for identifying robust analogs. Therefore separate models, one with EOF basis functions (HBA1) and a second model with Laplacian eignmap basis functions (HBA2), were implemented. Approximately $82 \%$ of the variation in the SST data was accounted for by the first 16 EOFs (i.e. $n_\alpha=16$). Laplacian eigenmaps are calculated through an eigenvector decomposition of a Laplacian matrix, whose construction involves an adjacency matrix formed through either a kernel or a nearest neighbor approach \citep[e.g.,][]{laplacianEigen}. We implemented the nearest neighbor approach, with $n_\alpha=16$ again, by sampling the number of neighbors as a parameter in the MCMC sampler over the following grid: $\{6+3 \times d : d=0,\dotsm\, 10 \}$. 

We used a value of $10^{-6}$ for the $\epsilon$ parameter in (\ref{processModel}); the model did not seem overly sensitive to this choice. For $q$ and $m$, we assigned priors $q \sim \text{DU}(30,60)$ and $m \sim \text{DU}(1,15)$, respectively. The kernel and process error parameters are given inverse-gamma priors:  $\theta_1 \sim \text{IG}(2.02,0.102)$ (which is only moderately informative in comparison to the small scale of the Gaussian kernel in (\ref{kern})) and $\sigma^2_\eta \sim \text{IG}(.001,.001)$.  All models were run for 20,000 iterations with the first 2,000 considered burn-in.

\section*{Results}

Prediction skill of the HBA and PST models was evaluated through calculation of the MSPE, defined as the mean of the squared differences between the posterior predicted means and the observed counts averaged across all spatial locations. The correlation between the observed counts and the mean of the posterior predictions was also used to evaluate the forecasting models, as is often considered for spatio-temporal prediction \citep[e.g.,][]{wilks2006}.  As displayed in Table \ref{table_2}, the HBA model out-performed the PST model in both 2009 and 2014, in that the HBA models had higher correlations and lower MSPE values for the two forecasted years. For 1999 and 2009, the EOF based analog model (HBA1) produced the most accurate results and the Laplacian eigenmaps model (HBA2) out-performed the EOF model in 2014. The correlation and MSPE for the hindcast appear to align with the results for the two forecasted years. We applied the model to several other hold-out years (not shown here) and found similar results, with the HBA always performing as well or better than the PST model and with the HBA1 generally, but not always, outperforming HBA2. 

\begin{table}[htb]
\begin{framed}
\captionsetup{font=footnotesize}
\begin{tabular}{ccccccccc} 
& \multicolumn{2}{c}{1999} & &  \multicolumn{2}{c}{2009} & &    \multicolumn{2}{c}{2014}    \\
Model & MSPE  & Corr & & MSPE &  Corr & &  MSPE & Corr  \\
\hline
HBA1& 58.822 & 83.031\% & &  63.056 & 70.307\%  & & 59.694 & 78.699\%   \\ 
HBA2 & 62.575 & 82.856\%  & & 70.085 & 66.808\%  && 57.799 & 79.446\%    \\ 
PST & - & -  & &  73.435 & 66.103\% & & 69.975 & 77.780\%   \\ 
\end{tabular}
\caption{Results based on the posterior predictive distribution for the two HBA models, and the PST model. Models are compared via mean squared prediction error (MSPE) and correlation (Corr) of the forecasted values with observed values. The two HBA models are implemented across 3 hold-out years, while the PST model is only evaluated for 2009 and 2014. }
\label{table_2}
\end{framed}
\end{table}

\begin{figure}[H]
\centering
\captionsetup{font=footnotesize}
\includegraphics[width=14cm,height=13.5cm]{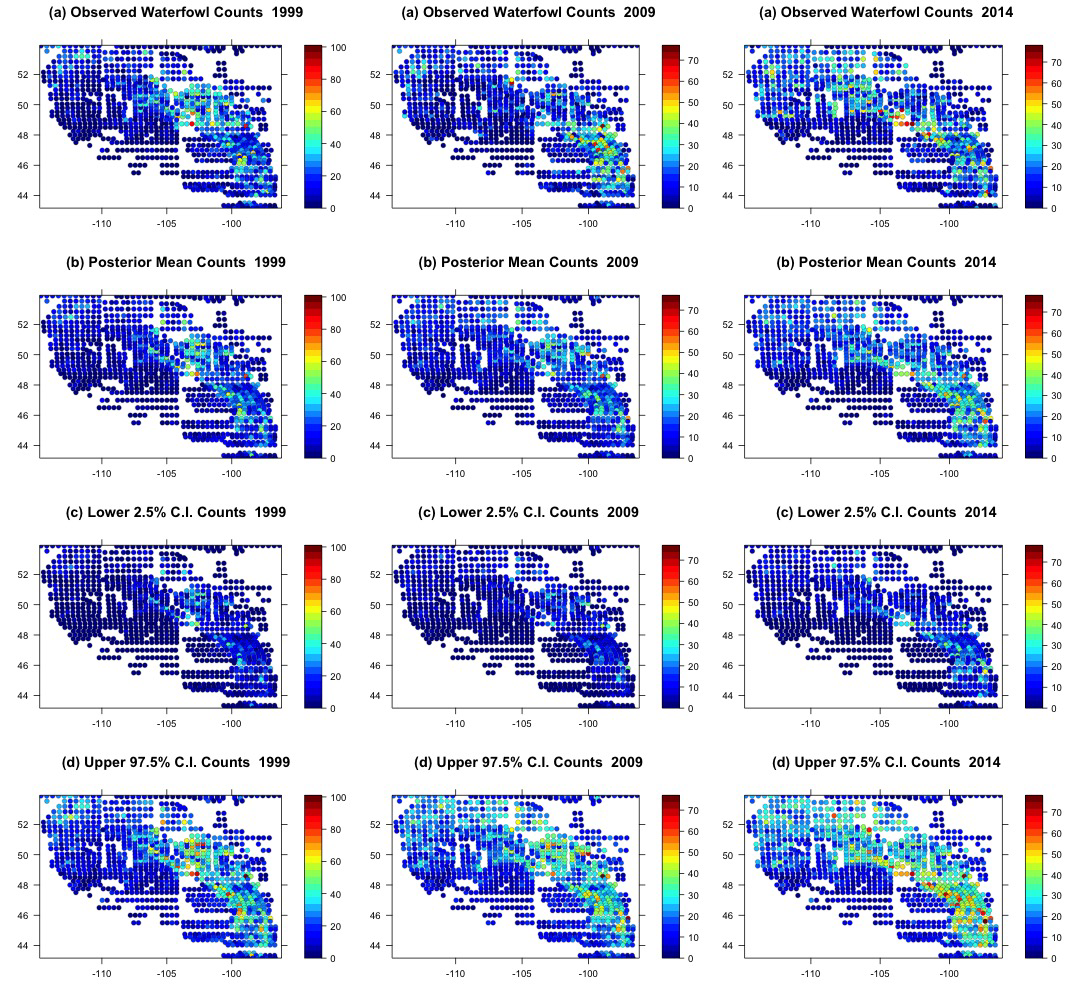}
\caption{Summary of the posterior predictive results for the HBA1 model. (a) Observed waterfowl counts for 1999, 2009, and 2014 (left to right), (b) means of the posterior predictive distribution for each year, (c)  lower 2.5th percentile from the posterior predictive distribution, and (d) upper 2.5th percentile form the posterior predictive distribution for each year.}  
\label{Figure_2}
\end{figure}

Figure \ref{Figure_2} shows the hindcast and prediction maps (observed, forecasted mean, site specific lower 2.5th, and upper 2.5th percentiles) from the posterior prediction distribution. The 1999 hindcast appears to correctly predict the pattern of mallards settling more heavily in the northern region of the domain. The forecasted maps for the two out-of sample periods (2009, 2014) also appear to capture the overall pattern of the observed counts. Close examination of the uncertainty maps show that a majority of the observed counts appear to fall within the displayed 95\% credible intervals. Overall, these spatial maps provide resource managers with accurate forecasts and informative intervals, thus allowing them to manage for various scenarios. It is important to note that, to our knowledge, no existing spatio-temporal nonlinear analog methods produce model based quantification of forecast uncertainty. 

\section*{Discussion}
Overall, many of the aspects of analog forecasting that originally made it appealing to ecologists are retained by the HBA model. The model has relatively few parameters, and performs well with data from a relatively short temporal span. Unlike other analog forecasting methods, the HBA allows users to properly quantify uncertainty in a rigorous framework. With the growing number of high-dimensional spatial-temporal ecological datasets, analog forecasting in a hierarchical framework can provide ecologists with a rich framework for making accurate forecasts with principled uncertainty quantification. The count-based spatio-temporal hierarchical Bayesian analog model methodology developed here was successful at forecasting waterfowl settling counts across multiple years. For example, the model correctly forecasted how the waterfowl settled more consistently in the northern half of the region of interest in 1999 despite the lack of correlation with patterns from the previous year. Further, there is a potential scientific explanation for this pattern. Poor habitat conditions  due to drought \citep[e.g.,][]{wu2013hierarchical}, possibly linked to the tropical Pacific La Ni\~na anomaly \citep[e.g,][]{philander1990,hoerling1997nino}, could help explain why many waterfowl overflew the southern region in 1999 \citep[e.g,][]{hansen1964emigration,sorenson1998potential}. 

%In comparison to a Poisson Bayesian hierarchical model with a latent dynamical spatio-temporal process (e.g., the PST model), the analog model consistently produced more accurate forecasts. While the linear PST model produced reasonable forecasts, the flexibility of the nonlinear HBA model produced more accurate forecasts.

Due to the preponderance of zeros present in the waterfowl data, the assumption of equidispersion implicit in the data model (i.e., equation (\ref{data_model})) is likely violated. \cite{wu2013hierarchical} attempted to deal with the potential underdispersion in such data by using a Conway-Maxwell Poisson data model. Here we deal with this potential problem by using NMF to reduce the overall underdispersion. However, a more rigorous way to deal with the underdispersion is to use a zero-inflated Poisson (ZIP) data model \citep[e.g.,][]{wikle2003climatological,ver2007space}. Importantly, any of the various methods throughout the literature that account for underdispersion could be integrated into the presented model by adjusting the data model.

By placing analog forecasting within a hierarchical Bayesian paradigm, there are a multitude of ways in which the methodology could be extended. It should be noted that differences in the forecasts between the HBA1 and HBA2 model can be attributed to a difference in the selection of analogs. This suggests that allowing the model to simultaneously consider multiple types of basis functions is an obvious extension of the model. Through the use of a mixture model, one could potentially jointly model two or more types of basis functions. Such an approach may be useful for forecasting seasonal or yearly settling patterns that are influenced both linearly and nonlinearly by some high-dimensional variable. 

%The dimension reduction methods presented here allow the HBA model to accommodate large datasets with relative ease. Although we found EOFs and Laplacian eigenmaps to be the most successful basis functions for the forcing variable, some of the other available nonlinear dimension reduction techniques \citep[e.g.,][]{lee2007nonlinear} may prove more useful for other applications such as modeling count data from radio-tracking animals.  Furthermore, the flexibility of the hierarchical modeling framework makes it relatively easy to incorporate other models that account for spatial or temporal dependence. For example, the process model could be extended to a mixture model between the analog model and a linear spatio-temporal dynamical model \citep[e.g., see ][]{wikle2015modern}.  One could also account for additional spatial dependence within the process model by including a more complicated covariance structure. If there was evidence of spatial or temporal structure in the response, the data model could also easily be extended to include random effect terms for these dependences. 
%
%Finally, due to the rapid advancement of telemetry technology, producing spatial-temporal models that can handle large data structures will be increasingly necessary for ecological research. The model presented here could be adapted to these or other types of data as well, by using different basis functions for the response (along with changing the distributions on the data and process model).

Count data in ecology are ubiquitous and the model we developed is an ideal alternative to currently available quantitative methods. Ecologists routinely collect count data through visual surveys, such as the waterfowl dataset used herein, or through use of other remote technologies.  For example, rapid advancement of radio-tracking technology \citep[e.g,][]{kays2015terrestrial} and remote-sensed cameras \citep[e.g,][]{he2016visual} have transformed the way ecologists collect count data.  These widely used technologies have also changed the type of data obtained both in terms of amount and structure of resulting data.  In particular, these technologies result in large data structures with spatial and temporal dependencies and our model provides an appropriate way to address these complexities while quantifying uncertainty in a rigorous manner.  Often, these count data are used by ecologists to assess settling patterns, habitat relationships, or impacts of weather conditions and predict future states.  For example, migration routes of terrestrial mammals are imperiled \citep[e.g,][]{berger2014moving} and there is much effort to identify and predict use of important migration corridors.  However, timing and use of
migration corridors is affected by weather and other factors such as human disturbance.  Our model provides an alternative to model and project use of these important areas while revealing factors affecting
their use.  Such results would have important policy decisions in wildlife management.  Thus, we envision numerous applications of this model and its extensions.

\section*{Acknowledgments}
This work was supported by the School of Natural Resources at the University of Missouri, the Missouri Department of Conservation, and was partially supported by the U.S. National Science Foundation (NSF) and the U.S. Census Bureau under NSF grant SES?1132031, funded through the NSF-Census Research Network (NCRN) program.

\section*{Data accessibility}
The raw indicated pair counts (for the waterfowl settling pattern data) are publicly available through the FWS Division of Migratory Management (\url{https://migbirdapps.fws.gov/}). The monthly Sea Surface Temperature (SST) data are publicly available from the National Oceanic and Atmospheric Administration (NOAA) Extended Reconstruction Sea Surface Temperature (ERSST) data (\url{http://www.esrl.noaa.gov/psd/}).

\newpage
\bibliography{reference}

\begin{thebibliography}{49}
\newcommand{\enquote}[1]{``#1''}
\expandafter\ifx\csname natexlab\endcsname\relax\def\natexlab#1{#1}\fi

\bibitem[{Arora et~al.(2013)Arora, Little, and McSharry}]{econGNP}
Arora, S., Little, M., and McSharry, P. (2013), \enquote{Nonlinear and
  nonparametric modeling approaches for probabilistic forecasting of the US
  gross national product,} \textit{Studies in Nonlinear Dynamics and
  Econometrics}, 17, 395--420.

\bibitem[{Badea(2008)}]{Badea}
Badea, L. (2008), \enquote{Extracting gene expression profiles common to colon
  and pancreatice adenocarinoma using simultaneous nonnegative matrix
  factorization,} \textit{Pacific Symposium of Biocomputing}, 267--278.

\bibitem[{Becker(2015)}]{becker2015search}
Becker, P.~H. (2015), \enquote{In search of the gap: temporal and spatial
  dynamics of settling in natal common tern recruits,} \textit{Behavioral
  Ecology and Sociobiology}, 69, 1415--1427.

\bibitem[{Belkin and Niyogi(2001)}]{laplacianEigen}
Belkin, M. and Niyogi, P. (2001), \enquote{Laplacian Eigenmaps and Spectral
  Techniques for Embedding and Clustering,} \textit{MIPS}, 14.

\bibitem[{Berger and Cain(2014)}]{berger2014moving}
Berger, J. and Cain, S.~L. (2014), \enquote{Moving beyond science to protect a
  mammalian migration corridor,} \textit{Conservation biology}, 28, 1142--1150.

\bibitem[{Berry et~al.(2007)Berry, Browne, Langville, and R.J.}]{Berry2007}
Berry, M., Browne, M., Langville, A., and R.J., P. V.~P. (2007),
  \enquote{Algorithms and applications for approximate nonnegative matrix
  factorization,} \textit{Compuational Statistics and Data Analysis}, 52,
  155--173.

\bibitem[{Boutsidis and Gallopoulos(2008)}]{Boutsidis}
Boutsidis, C. and Gallopoulos, E. (2008), \enquote{SVD based initialization: A
  head start for nonnegative matrix factorization,} \textit{Pattern
  Recognition}, 41, 1350--1362.

\bibitem[{Cangelosi and Hooten(2009)}]{cangelosi}
Cangelosi, A. and Hooten, M. (2009), \enquote{Models for bounded systems with
  continuous dynamics,} \textit{Biometrics}, 65, 850--856.

\bibitem[{Clark et~al.(2001)Clark, Carpenter, Barber, Collins, Dobson, Foley,
  Lodge, Pascual, Pielke~Jr, Pizer, et~al.}]{clark2001ecological}
Clark, J.~S., Carpenter, S.~R., Barber, M., Collins, S., Dobson, A., Foley,
  J.~A., Lodge, D.~M., Pascual, M., Pielke~Jr, R., Pizer, W., et~al. (2001),
  \enquote{Ecological forecasts: an emerging imperative,} \textit{science},
  293, 657--660.

\bibitem[{Coifman and Lafon(2006)}]{diffusionMaps}
Coifman, R. and Lafon, S. (2006), \enquote{Diffusion maps,} \textit{Applied
  computational Harmonic Analysis}, 21, 5--30.

\bibitem[{Conroy et~al.(2011)Conroy, Runge, Nichols, Stodola, and
  Cooper}]{conroy2011conservation}
Conroy, M.~J., Runge, M.~C., Nichols, J.~D., Stodola, K.~W., and Cooper, R.~J.
  (2011), \enquote{Conservation in the face of climate change: the roles of
  alternative models, monitoring, and adaptation in confronting and reducing
  uncertainty,} \textit{Biological Conservation}, 144, 1204--1213.

\bibitem[{Cressie et~al.(2009)Cressie, Calder, Clark, Hoef, and
  Wikle}]{cressie2009}
Cressie, N., Calder, C.~A., Clark, J.~S., Hoef, J. M.~V., and Wikle, C.~K.
  (2009), \enquote{Accounting for uncertainty in ecological analysis: the
  strengths and limitations of hierarchical statistical modeling,}
  \textit{Ecological Applications}, 19, 553--570.

\bibitem[{Cressie and Wikle(2011)}]{CandW2011}
Cressie, N. and Wikle, C. (2011), \textit{Statistics for Spatio-Temporal Data},
  New York: John Wiley \& Sons.

\bibitem[{Ellison(2004)}]{ellison2004bayesian}
Ellison, A.~M. (2004), \enquote{Bayesian inference in ecology,} \textit{Ecology
  letters}, 7, 509--520.

\bibitem[{Ellison and Gotelli(2004)}]{ellison2004primer}
Ellison, G.~N. and Gotelli, N. (2004), \enquote{A primer of ecological
  statistics,} \textit{Sinauer, Sunderland, Massachusetts, USA}.

\bibitem[{Feldman et~al.(2015)Feldman, Anderson, Howerter, and
  Murray}]{feldman2015does}
Feldman, R.~E., Anderson, M.~G., Howerter, D.~W., and Murray, D.~L. (2015),
  \enquote{Where does environmental stochasticity most influence population
  dynamics? An assessment along a regional core-periphery gradient for prairie
  breeding ducks,} \textit{Global Ecology and Biogeography}, 24, 896--904.

\bibitem[{Hansen and McKnight(1964)}]{hansen1964emigration}
Hansen, H.~A. and McKnight, D.~E. (1964), \enquote{Emigration of
  drought-displaced ducks to the Arctic,} in \textit{Transactions of the North
  American Wildlife and Natural Resources Conference}, vol.~29, pp. 119--127.

\bibitem[{Hastie et~al.(2013)Hastie, Tibshirani, and Friedman}]{HTF}
Hastie, T., Tibshirani, R., and Friedman, J. (2013), \textit{The Elements of
  Statistical Learning Data mining, Inference, and Prediction}, New York:
  Springer.

\bibitem[{He et~al.(2016)He, Kays, Zhang, Ning, Huang, Han, Millspaugh,
  Forrester, and McShea}]{he2016visual}
He, Z., Kays, R., Zhang, Z., Ning, G., Huang, C., Han, T.~X., Millspaugh, J.,
  Forrester, T., and McShea, W. (2016), \enquote{Visual Informatics Tools for
  Supporting Large-Scale Collaborative Wildlife Monitoring with Citizen
  Scientists,} \textit{IEEE Circuits and Systems Magazine}, 16, 73--86.

\bibitem[{Herter(2012)}]{herter2012bird}
Herter, D. (2012), \enquote{Bird Migration in the Arctic: A Review,}
  \textit{Bird Migration: Physiology and Ecophysiology}, 22.

\bibitem[{Hoerling et~al.(1997)Hoerling, Kumar, and Zhong}]{hoerling1997nino}
Hoerling, M.~P., Kumar, A., and Zhong, M. (1997), \enquote{El Ni{\~n}o, La
  Ni{\~n}a, and the nonlinearity of their teleconnections,} \textit{Journal of
  Climate}, 10, 1769--1786.

\bibitem[{Johnson and Grier(1988)}]{johnson1988determinants}
Johnson, D.~H. and Grier, J.~W. (1988), \enquote{Determinants of breeding
  distributions of ducks,} \textit{Wildlife Monographs}, 3--37.

\bibitem[{Karanth et~al.(2014)Karanth, Nichols, Sauer, Hines, and
  Yackulic}]{karanth2014latitudinal}
Karanth, K.~K., Nichols, J.~D., Sauer, J.~R., Hines, J.~E., and Yackulic, C.~B.
  (2014), \enquote{Latitudinal gradients in North American avian species
  richness, turnover rates and extinction probabilities,} \textit{Ecography},
  37, 626--636.

\bibitem[{Kays et~al.(2015)Kays, Crofoot, Jetz, and
  Wikelski}]{kays2015terrestrial}
Kays, R., Crofoot, M.~C., Jetz, W., and Wikelski, M. (2015),
  \enquote{Terrestrial animal tracking as an eye on life and planet,}
  \textit{Science}, 348, aaa2478.

\bibitem[{LeBrun et~al.(2016)LeBrun, Thogmartin, Thompson, Dijak, and
  Millspaugh}]{lebrun2016assessing}
LeBrun, J.~J., Thogmartin, W.~E., Thompson, F.~R., Dijak, W.~D., and
  Millspaugh, J.~J. (2016), \enquote{Assessing the sensitivity of avian species
  abundance to land cover and climate,} \textit{Ecosphere}, 7.

\bibitem[{Lee and Seung(2001)}]{nmf2001}
Lee, D. and Seung, H. (2001), \enquote{Algorithms for non-negative matrix
  factorization,} in \textit{Advances in neural information processing
  systems}, eds. Leen, T.~K., Dietterich, T.~G., and Tresp, V., MIT Press, pp.
  556--562.

\bibitem[{Lorenz(1969)}]{lorenz1969atmospheric}
Lorenz, E.~N. (1969), \enquote{Atmospheric predictability as revealed by
  naturally occurring analogues,} \textit{Journal of the Atmospheric sciences},
  26, 636--646.

\bibitem[{McDermott and Wikle(2016)}]{analog2016}
McDermott, P. and Wikle, C. (2016), \enquote{A model based approach for analog
  spatio-temporal dynamic forecasting,} \textit{Environmetrics}.

\bibitem[{Oliver and Roy(2015)}]{oliver2015pitfalls}
Oliver, T.~H. and Roy, D.~B. (2015), \enquote{The pitfalls of ecological
  forecasting,} \textit{Biological Journal of the Linnean Society}, 115,
  767--778.

\bibitem[{Perretti et~al.(2013)Perretti, Munch, and
  Sugihara}]{perretti2013model}
Perretti, C.~T., Munch, S.~B., and Sugihara, G. (2013), \enquote{Model-free
  forecasting outperforms the correct mechanistic model for simulated and
  experimental data,} \textit{Proceedings of the National Academy of Sciences},
  110, 5253--5257.

\bibitem[{Philander(1990)}]{philander1990}
Philander, S. (1990), \textit{El Ni\~no, La Ni\~na, and the Southern
  Oscillation}, Academic Press, San Diego.

\bibitem[{Robert and Casella(2004)}]{robert2004monte}
Robert, C. and Casella, G. (2004), \enquote{Monte Carlo Statistical Methods
  Springer-Verlag,} \textit{New York}.

\bibitem[{Roweis and Saul(2000)}]{lle}
Roweis, S. and Saul, L. (2000), \enquote{Nonlinear dimensionality reduction by
  locally linear embedding,} \textit{Science}, 290, 2232--2326.

\bibitem[{Royle and Dorazio(2008)}]{royle2008hierarchical}
Royle, J.~A. and Dorazio, R.~M. (2008), \textit{Hierarchical modeling and
  inference in ecology: the analysis of data from populations, metapopulations
  and communities}, Academic Press.

\bibitem[{Scholkopf et~al.(1998)Scholkopf, Smola, and Muller}]{kernelPCA}
Scholkopf, B., Smola, A., and Muller, K. (1998), \enquote{Nonlinear component
  analysis as a kernel eigenvalue problem,} \textit{Neural computation}, 10,
  1299--1319.

\bibitem[{Sorenson et~al.(1998)Sorenson, Goldberg, Root, and
  Anderson}]{sorenson1998potential}
Sorenson, L.~G., Goldberg, R., Root, T.~L., and Anderson, M.~G. (1998),
  \enquote{Potential effects of global warming on waterfowl populations
  breeding in the northern Great Plains,} \textit{Climatic change}, 40,
  343--369.

\bibitem[{Sugihara and May(1990)}]{sug}
Sugihara, G. and May, R. (1990), \enquote{Nonlinear forecasting as a way of
  distinguishing from measurement error in time series,} \textit{Nature}, 344,
  734--741.

\bibitem[{Sugihara et~al.(2012)Sugihara, May, Ye, Hsieh, Deyle, Fogarty, and
  Munch}]{sug2012}
Sugihara, G., May, R., Ye, H., Hsieh, C., Deyle, E., Fogarty, M., and Munch, S.
  (2012), \enquote{Detecting causality in complex ecosystems,}
  \textit{Science}, 338.

\bibitem[{Takens(1981)}]{Takens}
Takens, F. (1981), \enquote{Detecting strange attractors in turbulence,}
  \textit{Dynamical Systems and Turbulence}, 898, 366--381.

\bibitem[{Tippett and DelSole(2013)}]{Tippett}
Tippett, M. and DelSole, T. (2013), \enquote{Constructed analogues and linear
  regression,} \textit{Monthly Weather Review}, 141, 2519--2525.

\bibitem[{Ver~Hoef and Jansen(2007)}]{ver2007space}
Ver~Hoef, J.~M. and Jansen, J.~K. (2007), \enquote{Space---time zero-inflated
  count models of Harbor seals,} \textit{Environmetrics}, 18, 697--712.

\bibitem[{Ward et~al.(2014)Ward, Holmes, Thorson, and
  Collen}]{ward2014complexity}
Ward, E.~J., Holmes, E.~E., Thorson, J.~T., and Collen, B. (2014),
  \enquote{Complexity is costly: a meta-analysis of parametric and
  non-parametric methods for short-term population forecasting,}
  \textit{Oikos}, 123, 652--661.

\bibitem[{Wikle(2003)}]{wikle2003}
Wikle, C. (2003), \enquote{Hierarcical Bayesian Models for predicting the
  spread of Ecological Processes,} \textit{Ecology}, 84, 1382--1394.

\bibitem[{Wikle(2015)}]{wikle2015modern}
--- (2015), \enquote{Modern perspectives on statistics for spatio-temporal
  data,} \textit{Wiley Interdisciplinary Reviews: Computational Statistics}, 7,
  86--98.

\bibitem[{Wikle and Anderson(2003)}]{wikle2003climatological}
Wikle, C.~K. and Anderson, C.~J. (2003), \enquote{Climatological analysis of
  tornado report counts using a hierarchical Bayesian spatiotemporal model,}
  \textit{Journal of Geophysical Research: Atmospheres}, 108.

\bibitem[{Wilks(2006)}]{wilks2006}
Wilks, D.~S. (2006), \textit{Statistical Methods in Atmospheric Sciences}, vol.
  100, Academic press.

\bibitem[{Wilks(2011)}]{wilks2011statistical}
--- (2011), \textit{Statistical methods in the atmospheric sciences}, vol. 100,
  Academic press.

\bibitem[{Wu et~al.(2013)Wu, Holan, and Wikle}]{wu2013hierarchical}
Wu, G., Holan, S.~H., and Wikle, C.~K. (2013), \enquote{Hierarchical Bayesian
  spatio-temporal Conway--Maxwell Poisson models with dynamic dispersion,}
  \textit{Journal of Agricultural, Biological, and Environmental Statistics},
  18, 335--356.

\bibitem[{Zhao and Giannakis(2014)}]{Zhao}
Zhao, Z. and Giannakis, D. (2014), \enquote{Analog forecasting with
  dynamics-adapted kernels,} \textit{Nonlinearity}, in review.

\end{thebibliography}
\newpage
\section*{Appendix A: Markov chain Monte Carlo Algorithm}
The following details the MCMC algorithm used to implement the HBA model outlined above. Let $\ell=1,\dots, L$ represent the current iteration. 
\begin{enumerate}
\item Sample $\beta_{j,t}^{(\ell)}$ using componentwise Metropolis-Hastings updates.

\vspace{.5cm}

Let $\widetilde{\mbv{\Theta}}^{(\ell-1)}=\{m^{(\ell-1)},q^{(\ell-1)},\theta_1^{(\ell-1)},\sigma_\eta^{2(\ell-1)}\}$. Generate a proposal value from $\text{log}(\beta_{j,t}^*) \sim \text{Gau}( \text{log}(\beta_{j,t}^{(\ell-1)}),\zeta_{j,t} )$ (where $\zeta_{j,t}$ is a tuning parameter) and calculate the following vectors:\\
$\mbv{\beta}_t^{0}=(\beta_{1,t}^{(\ell)},\dots,\beta_{j-1,t}^{(\ell)},\beta_{j,t}^{(\ell-1)},\beta_{j+1,t}^{(\ell-1)},\dots,\beta_{n_\beta,t}^{(\ell-1)})'$\\
$\mbv{\beta}_t^*=(\beta_{1,t}^{(\ell)},\dots,\beta_{j-1,t}^{(\ell)},\beta_{j,t}^*,\beta_{j+1,t}^{(\ell-1)},\dots,\beta_{n_\beta,t}^{(\ell-1)})'$\\
$\mbv{\beta}_j^0=(\beta_{j,1}^{(\ell)},\dots,\beta_{j,t-1}^{(\ell)},\beta_{j,t}^{(\ell-1)},\beta_{j,t+1}^{(\ell-1)},\dots,\beta_{j,T}^{(\ell-1)})'$\\
$\mbv{\beta}_j^*=(\beta_{j,1}^{(\ell)},\dots,\beta_{j,t-1}^{(\ell)},\beta_{j,t}^*,\beta_{j,t+1}^{(\ell-1)},\dots,\beta_{j,T}^{(\ell-1)})'$.

\vspace{.5cm}

Next, calculate the following Metropolis-Hastings ratio:

\vspace{.5cm}

$R(\beta_{j,t}^*,\beta_{j,t}^{(\ell-1)})=\frac{\big[\mbv{Y}_t \mid \mbv{\beta}_t^*, \mbv{\Psi}\big] \ \mathlarger{\prod\limits_{t=1}^T} \ \big[ \mbv{\beta}_{j,t}^*\mid \mbv{\beta}_{j,-t}^*, \widetilde{\mbv{\Theta}}^{(\ell-1)} \big] }{\big[\mbv{Y}_t \mid \mbv{\beta}_t^{0}, \mbv{\Psi}\big] \ \mathlarger{\prod\limits_{t=1}^T} \ \big[ \mbv{\beta}_{j,t}^0\mid \mbv{\beta}_{j,-t}^0, \widetilde{\mbv{\Theta}}^{(\ell-1)}  \big]}$. 

\vspace{.35cm}

Set $\beta_{j,t}^{(\ell)}=\beta_{j,t}^*$ with probability $\text{min}\{1,R(\beta_{j,t}^*,\beta_{j,t}^{(\ell-1)})\}$; otherwise $\beta_{j,t}^{(\ell)}=\beta_{j,t}^{(\ell-1)}$. Repeat for $j=1,\dots, n_\beta$ and $t=1,\dots, T$. This derivation assumes each $\beta_{j,t}$ are sampled in the same order each iteration, in practice the order could be random and change each iteration.
\item Sample $m$ by performing inverse transform sampling with the discrete grid, $\{m_i^* : i=1,\dots,(m_{max}-m_{min},+1)\}$. Evaluate the following probability: 

\vspace{.5cm}

$p_i=\mathlarger{\prod\limits_{t=1}^T} \ \big[ \mbv{\beta}_t^{(\ell)} \mid \mbv{\beta}_{-t}^{(\ell)},m_i^*,q^{(\ell-1)},\theta_1^{(\ell-1)},\sigma_\eta^{2(\ell-1)}   \big]$,

\vspace{.5cm}

for each $m_i^*$. Calculate the C.D.F. by normalizing  each $p_i$ (i.e., $\tilde{p}_i=\frac{p_i}{\sum_i p_i}$ ) and use inverse transform sampling to sample $m^{(\ell)}$.
\item Sample $q$ by performing inverse transform sampling with the discrete grid, $\{q_k^* : k=1,\dots,(q_{max}-q_{min},+1)\}$. Evaluate the following probability: 

\vspace{.5cm}

$p_k=\mathlarger{\prod\limits_{t=1}^T} \ \big[ \mbv{\beta}_t^{(\ell)} \mid \mbv{\beta}_{-t}^{(\ell)},m^{(\ell)},q_k^*,\theta_1^{(\ell-1)},\sigma_\eta^{2(\ell-1)}   \big]$,

\vspace{.5cm}

for each $q_k^*$. Calculate the C.D.F. by normalizing  each $p_k$ and use inverse transform sampling to sample $q^{(\ell)}$.

\item Sample $\theta_1^{(\ell)}$ with a Metropolis-Hastings step. Generate a proposal value from $\text{log}(\theta^*_1)\sim \text{Gau}(\text{log}(\theta_1^{(\ell-1)}),\sigma_{\theta_1}^2) $ and calculate the following Metropolis-Hastings ratio:

\vspace{.5cm}

$R(\theta^*_1,\theta_1^{(\ell-1)})=\frac{\mathlarger{\prod\limits_{t=1}^T} \ \big[ \mbv{\beta}_t^{(\ell)} \mid \mbv{\beta}_{-t}^{(\ell)},m^{(\ell)},q_k^{(\ell)},\theta_1^*,\sigma_\eta^{2(\ell-1)}   \big] \ \big[ \theta_1^* \mid a_1,b_1 \big]}{\mathlarger{\prod\limits_{t=1}^T} \ \big[ \mbv{\beta}_t^{(\ell)} \mid \mbv{\beta}_{-t}^{(\ell)},m^{(\ell)},q_k^{(\ell)},\theta_1^{(\ell-1)},\sigma_\eta^{2(\ell-1)}   \big] \ \big[ \theta_1^{(\ell-1)} \mid a_1,b_1 \big]}$.

\vspace{.35cm}

Set $\theta_1^{(\ell)}=\theta_1^*$ with probability $\text{min}\{1,R(\theta^*_1,\theta_1^{(\ell-1)})\}$; otherwise $\theta_1^{(\ell)}=\theta_1^{(\ell-1)}$.

\item Sample $\sigma_\eta^{2(\ell)}$ with a Metropolis-Hastings step. Generate a proposal value from $\text{log}(\sigma_\eta^{2*})\sim \text{Gau}(\text{log}(\sigma_\eta^{2(\ell-1)}),\sigma_{\sigma^2_\eta}^2) $ and calculate the following Metropolis-Hastings ratio:

\vspace{.5cm}

$R(\sigma_\eta^{2*},\sigma_\eta^{2(\ell-1)})=\frac{\mathlarger{\prod\limits_{t=1}^T} \ \big[ \mbv{\beta}_t^{(\ell)} \mid \mbv{\beta}_{-t}^{(\ell)},m^{(\ell)},q_k^{(\ell)},\theta_1^{(\ell)},\sigma_\eta^{2*}   \big] \ \big[ \sigma_\eta^{2*} \mid a_2,b_2 \big]}{\mathlarger{\prod\limits_{t=1}^T} \ \big[ \mbv{\beta}_t^{(\ell)} \mid \mbv{\beta}_{-t}^{(\ell)},m^{(\ell)},q_k^{(\ell)},\theta_1^{(\ell)},\sigma_\eta^{2(\ell-1)}   \big] \ \big[ \sigma_\eta^{2(\ell-1)} \mid a_2,b_2 \big]}$.

\vspace{.35cm}

Set $\sigma_\eta^{2(\ell)}=\sigma_\eta^{2*}$ with probability $\text{min}\{1,R(\sigma_\eta^{2*},\sigma_\eta^{2(\ell-1)})\}$; otherwise $\sigma_\eta^{2(\ell)}=\sigma_\eta^{2(\ell-1)}$.

\end{enumerate}

\section*{Appendix B: Procrustes Distance}
Suppose we have a target object matrix $\bE$ and a comparison object matrix $\bF$. It is assumed that $\bE$ and $\bF$ have the same dimension. To compare the two objects the comparison object is superimposed onto the target object through scaling, rotation, and translation. This transformation is carried out by using the scaling parameter $\theta_2$ and the rotation matrix $\bR$. The  Procrustes distance between $\bE$ and $\bF$ is defined as: 

\vspace{.5cm}

$d(\bE,\bF; \theta_2)= ||\bE -\theta_2\bF \bR ||_{F}$ ,

\vspace{.5cm}

where the ``F" subscript denotes the Frobenius matrix norm. To calculate $\bR$ we need  to center $\bE$ and $\bF$ by their respective column means to create $\widetilde{\bE}$ and $\widetilde{\bF}$. Next, we calculate the singular value decomposition of $\widetilde{\bE}\widetilde{\bF}'=\bU\bD\bV'$ and let $\bR=\bU\bV'$. The positive scaling parameter is set such that $\theta_2=tr(\bD)/ ||\bF||^2_{F}$. 

\end{document}